\documentclass[12pt]{article}

\pdfoutput=1
\usepackage{color}
\usepackage{epsfig,palatino}
\usepackage{mathrsfs}
\usepackage{lmodern} 
\usepackage[T1]{fontenc}
\usepackage[ansinew]{inputenc}
\usepackage{amsmath,amssymb,amsthm}
\usepackage{graphicx}
\usepackage{xcolor}
\definecolor{darkblue}{cmyk}{0.9,0.9,0,0}
\usepackage[colorlinks=true,linkcolor=darkblue,citecolor=darkblue,urlcolor=darkblue]{hyperref}
\usepackage{cite}
\usepackage{wasysym}
\usepackage{varioref}
\usepackage[english]{babel}
\usepackage{tikz}
\usepackage{booktabs}

\usepackage[font={small}]{caption}

\usepackage{microtype}

\newtheorem{theorem}{Theorem}

\newcommand{\td}[1][]{\mathrm{d}^{#1}}
\newcommand{\defas}{\mathrel{\mathop:}=}
\newcommand{\U}{\mathcal{U}}
\newcommand{\F}{\mathcal{F}}
\newcommand{\sdc}{\omega}
\newcommand{\dimeps}{\varepsilon}
\newcommand{\DIM}{d}
\newcommand{\ZZ}{\mathbb{Z}}
\newcommand{\RR}{\mathbb{R}}
\newcommand{\mzv}[1]{\zeta(#1)}
\newcommand{\mzveps}[1]{\hat{\zeta}(#1)}
\newcommand{\GnC}{GaC}
\newcommand{\LM}{\ell}

\newcommand{\Transpose}{\intercal}

\newcommand{\Azurite}{\href{https://bitbucket.org/yzhphy/azurite/}{\textsc{Azurite}}}
\newcommand{\FIRE}{\href{https://bitbucket.org/feynmanIntegrals/fire/}{\textsc{FIRE}}}

\begin{document}

\thispagestyle{empty}

\renewcommand{\thefootnote}{\fnsymbol{footnote}}
\setcounter{page}{1}
\setcounter{footnote}{0}
\setcounter{figure}{0}

%%%%%%%%%%%%%%%%%%%%%%%%%%%%%%%%%%%%%%%%%%%%%%%%%%%%%%%%%%%%%%%%%%%%%%%%%%%%%%%%%%%%%%%%%%%%%%%%%%%

\noindent

\hfill
\begin{minipage}[t]{35mm}
\begin{flushright}
UUITP-37/20 \\
\end{flushright}
\end{minipage}

\vspace{1.0cm}

\begin{center}
{\Large\textbf{\mathversion{bold}
Glue-and-Cut at Five Loops \\
}\par}

\vspace{1.0cm}

\textrm{Alessandro Georgoudis\textsuperscript{1,2,4}, Vasco Goncalves\textsuperscript{3,4}, Erik Panzer\textsuperscript{5}, Raul Pereira\textsuperscript{6}, Alexander V. Smirnov\textsuperscript{7,8}, Vladimir A. Smirnov\textsuperscript{8,9}}
\\ \vspace{1.2cm}
\footnotesize{\textit{
\textsuperscript{1}Laboratoire de physique de l'Ecole normale sup\'erieure, ENS, Universit\'e PSL, CNRS, Sorbonne Universit\'e, Universit\'e Paris-Diderot, Sorbonne Paris Cit\'e, 24 rue Lhomond, 75005 Paris, France\\
\textsuperscript{2}Department of Physics and Astronomy, Uppsala University
Box 516, SE-751 20 Uppsala, Sweden\\
\textsuperscript{3}Centro de F\'isica do Porto e Departamento de F\'isica e Astronomia, Faculdade de Ci\^encias da Universidade do Porto, Rua do Campo Alegre 687, 4169-007 Porto, Portugal\\
\textsuperscript{4}ICTP South American Institute for Fundamental Research, Instituto de F\'isica Te\'orica, Universidade Estadual Paulista, Rua Dr. Bento T. Ferraz 271, 01140-70, Sao Paulo, Brazil\\
\textsuperscript{5}Mathematical Institute, University of Oxford, OX2 6GG, Oxford, UK\\
\textsuperscript{6}School of Mathematics and Hamilton Mathematics Institute, Trinity College Dublin, Dublin, Ireland\\
\textsuperscript{7}Research Computing Center, Moscow State University,119992 Moscow, Russia\\
\textsuperscript{8}Moscow Center for Fundamental and Applied Mathematics, 119991 Moscow, Russia \\
\textsuperscript{9}Skobeltsyn Institute of Nuclear Physics of Moscow State University,119992 Moscow, Russia
}  
\vspace{4mm}
}

\end{center}

\begin{abstract}
	We compute $\dimeps$-expansions around 4 dimensions of a complete set of master integrals for momentum space five-loop massless propagator integrals in dimensional regularization, up to and including the first order with contributions of transcendental weight nine. Our method is the glue-and-cut technique from Baikov and Chetyrkin, which proves extremely effective in that it determines all expansion coefficients to this order in terms of recursively one-loop integrals and only one further integral. We observe that our results are compatible with conjectures that predict $\pi$-dependent contributions.
\end{abstract}
\noindent

\setcounter{page}{1}
\renewcommand{\thefootnote}{\arabic{footnote}}
\setcounter{footnote}{0}

\setcounter{tocdepth}{2}

\newpage

\parskip 5pt plus 1pt   \jot = 1.5ex

\newpage

\parskip 5pt plus 1pt   \jot = 1.5ex

\section{Introduction}

Despite ongoing efforts do develop alternative representations, practical calculations of scattering amplitudes in quantum field theories are still described mostly through Feynman diagrams. As one studies observables at higher orders in perturbation theory in this framework, one needs to evaluate a huge number of corresponding many-loop Feynman integrals. The predominant approaches exploit integration by parts (IBP) identities in momentum space \cite{Tkachov:calculability-4loop,Chetyrkin:1981qh} to express all required integrals as linear combinations of a much smaller number \cite{SmirnovPetukhov:Finite,LeePomeransky:CriticalPoints,BitounBognerKlausenPanzer:AnnihilatorsEuler} of basis integrals, called master integrals (MI). Although this reduction procedure may become challenging in practice, it is thus in principle sufficient to compute and tabulate only one such set of master integrals for a given problem.
Due to the presence of divergences in the integrals, calculations are performed in dimensional regularization in $\DIM=4-2\dimeps$ dimensions \cite{BolliniGiambiagi:DimReg,tHooftVeltman:RegularizationGaugeFields}, and the master integrals have to be expanded to a sufficiently high order in the $\dimeps$-expansion.

In this paper we focus on the evaluation of massless propagator-type integrals in momentum space, also known as ``p-integrals''. The importance of p-integrals is demonstrated by the success of their application to determine arbitrary renormalization group functions up to four loops \cite{Tkachov:calculability-4loop,Chetyrkin:1981qh}, in scalar $\phi^4$ theory up to 6 loops \cite{KNFCL:5loopPhi4,Kompaniets:2017yct}, and recently to five loops in QCD \cite{BaikovChetyrkinKuehn:5loopQCD,HerzogRuijlUedaVermaserenVogt:5loopYMferm}. Alternative effective routes to renormalization group functions include massive vacuum integrals \cite{RitbergenVermaserenLarin:4BetaQCD,Czakon:4loopQCDbeta,LutheMaierMarquardSchroeder:CompleteQCD5} and position-space techniques \cite{Schnetz:NumbersAndFunctions}. In this comparison, the coefficients of the $\dimeps$-expansions of p-integrals tend to be rather simple and are often expressible exactly in terms of Riemann zeta values. On the other hand, masslessness introduces infrared divergences, which have to be carefully separated from the sought-after ultraviolet singularities. This is achieved with the $R^\ast$-operation \cite{ChetyrkinSmirnov:Rcorrected,ChetyrkinTkachov:InfraredR,Chetyrkin:RRR}, which is now in a very mature state \cite{HerzogRuijl:Rstar,BeekveldtBorinskyHerzog:HopfR} and provides a technique that expresses $L$-loop renormalization group functions in terms of poles and finite terms of the $\dimeps$-expansions of p-integrals with only $L-1$ loops.

As a step towards an extension of the currently very good
understanding of the master p-integrals
\cite{Lee:2011jt,Baikov:2010hf,SmirnovTentyukov:FourLoopPropagatorsNumeric}
and their IBP reduction \cite{Ueda:2016yjm} at four loops, we compute
in this paper a complete set of master p-integrals at five loops,
which should be useful in future determinations of renormalization
group functions at the six-loop order. Our results are also
interesting from a structural point of view and give further evidence
for general patterns in the coefficients of $\dimeps$-expansions of
p-integrals, and therefore, about the coefficients of renormalization
group functions
\cite{Baikov:2019zmy,KotikovTeber:LKFzeta}. Concretely, we confirm
that finite five-loop p-integrals in $\DIM=4$ dimensions are linear
combinations of multiple zeta values of transcendental weight at most
nine, and that, in a well-defined sense, all $\pi$-dependent
contributions can be predicted by the $\pi$-free terms. We supply our
results for the $\dimeps$-expansions of our master integrals and other
results in the ancillary files that supplement this article, the use and content of each file is described in the \texttt{README} file.

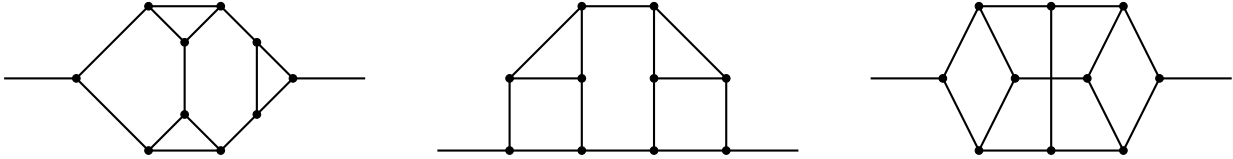
\begin{figure}
\begin{tikzpicture}[scale=0.48]
 \draw[thick] (0,0) -- (2,0)--(4,2)--(6,2)--(8,0)--(10,0);
 \draw[thick] (2,0) --(4,-2)--(6,-2)--(8,0);
 \draw[thick] (4,2) --(5,1)--(6,2);
 \draw[thick] (4,-2) --(5,-1)--(6,-2); 
 \draw[thick] (5,1) --(5,-1);
 \draw[thick] (7,-1)--(7,1);
 \draw[thick] (12,-2)--(22,-2);
 \draw[thick] (14,-2)--(14,0)--(16,2)-- (18,2)--(20,0)--(20,-2);
 \draw[thick] (16,-2)--(16,2);
 \draw[thick] (18,2)--(18,-2);
 \draw[thick] (14,0)--(16,0);
 \draw[thick] (18,0)--(20,0);
 \draw[thick] (24,0)--(26,0)--(27,2)--(31,2)--(32,0)--(34,0);
 \draw[thick] (26,0)--(27,-2)--(31,-2)--(32,0);
 \draw[thick] (27,2)--(28,0)--(27,-2);
 \draw[thick] (31,-2)--(30,0)--(31,2);
 \draw[thick] (28,0)--(30,0);
 \draw[thick] (29,-2)--(29,2);
 \filldraw[fill=black] (2,0) circle (3pt);
 \filldraw[fill=black] (4,2) circle (3pt);
 \filldraw[fill=black] (6,2) circle (3pt);
 \filldraw[fill=black] (8,0) circle (3pt);
 \filldraw[fill=black] (4,-2) circle (3pt);
 \filldraw[fill=black] (6,-2) circle (3pt);
 \filldraw[fill=black] (5,1) circle (3pt);
 \filldraw[fill=black] (5,-1) circle (3pt);
 \filldraw[fill=black] (7,1) circle (3pt);
 \filldraw[fill=black] (7,-1) circle (3pt);

 \filldraw[fill=black] (14,-2) circle (3pt);
 \filldraw[fill=black] (14,0) circle (3pt);
 \filldraw[fill=black] (16,2) circle (3pt);
 \filldraw[fill=black] (18,2) circle (3pt);
 \filldraw[fill=black] (20,0) circle (3pt);
 \filldraw[fill=black] (20,-2) circle (3pt);
 \filldraw[fill=black] (16,-2) circle (3pt);
 \filldraw[fill=black] (18,-2) circle (3pt);
 \filldraw[fill=black] (18,0) circle (3pt);
 \filldraw[fill=black] (16,0) circle (3pt);

 \filldraw[fill=black] (26,0) circle (3pt);
 \filldraw[fill=black] (27,2) circle (3pt);
 \filldraw[fill=black] (31,2) circle (3pt);
 \filldraw[fill=black] (32,0) circle (3pt);
 \filldraw[fill=black] (27,-2) circle (3pt);
 \filldraw[fill=black] (31,-2) circle (3pt);
 \filldraw[fill=black] (28,0) circle (3pt);
 \filldraw[fill=black] (30,0) circle (3pt);
 \filldraw[fill=black] (29,-2) circle (3pt);
 \filldraw[fill=black] (29,2) circle (3pt);
\end{tikzpicture}
	\caption{These are three examples of maximal sectors for five-loop p-integrals. The diagrams have only cubic vertices and $14$ internal propagators. The rightmost sector is an example of a non-planar integral.} \label{PInts}%
\end{figure}

All five-loop p-integrals can be obtained by contracting edges of one of 64 maximal sectors\footnote{
 A \emph{sector} encompasses all integrals in a family $P_k(a;\DIM)$ where the set $\{j\colon a_j=0\}$ is fixed. Since setting $a_j=0$ for an inverse propagator ($j<15$) corresponds to contraction of the associated edge in the underlying graph of the family, different families can share common subsectors (different graphs may become isomorphic after several contractions).}, by which we mean diagrams with only cubic vertices and 14 internal propagators, as illustrated in \autoref{PInts}.
Note that the planar graphs can also be viewed as propagator integrals in position-space, and we completely determined those in previous work \cite{Georgoudis:2018olj}. Our calculation presented here confirms these earlier findings for the planar p-integrals, but provides also the expansions for the non-planar integrals.

A range of methods to compute single-scale massless Feynman integrals has been developed over time \cite{KotikovTeber:MultiTechniques}. Relatively recent advances and applications include in particular dimensional recurrence relations \cite{Lee:2009dh,Lee:2011jt}, integration over Schwinger parameters \cite{Panzer:MasslessPropagators,Brown:TwoPoint,Panzer:2014caa,ManteuffelPanzerSchabinger:QuasiFinite} and graphical functions \cite{Schnetz:GraphicalFunctions}.

In order to determine all five-loop p-integrals we follow the algorithm introduced at four loops by Baikov and Chetyrkin \cite{Baikov:2010hf}, which combines the glue-and-cut ({\GnC}) symmetry \cite{Chetyrkin:1980pr} of massless p-integrals with IBP reductions to bootstrap the determination of master integrals. This elegant algorithm manages to get by without having to actually compute a lot of complicated integrals explicitly. Instead, it produces so many relations between the coefficients of the $\dimeps$-expansions of the p-integrals, that ultimately all get reduced to coefficients of only a small number of particularly simple integrals. Concretely, the constraints turned out to be so strong that the input of recursively one-loop integrals (see \autoref{Rol}), which are easily evaluated to all orders in $\dimeps$, was sufficient to determine all master integrals at four loops \cite{Baikov:2010hf}.

The bottleneck in this strategy resides mostly in the efficiency of IBP reduction routines, and only recent progress in that field has allowed us to finally apply the {\GnC} strategy at five loops and to fix all p-integrals at that order.
We find that it suffices to input the expansions of $21$ recursively one-loop master integrals, together with only one further integral. In fact, this single additional input datum can be chosen in form of a very simple product integral, see \autoref{Extra}. This demonstrates strikingly that the remarkable power of the {\GnC} method persists at the five loop order, circumventing completely the daunting direct evaluation of a large number of very complicated Feynman integrals.

The paper is organized as follows. In \autoref{review} we review the main ideas of the method and describe how we obtain the relations needed for bootstrapping the p-integrals. In \autoref{convergence} we describe an algorithm to determine the convergence of vacuum diagrams. In \autoref{reductions} we describe how we performed the IBP reductions needed while in \autoref{details} we describe how we generated the needed vacuum diagrams and how we constructed and solved the equations needed for the bootstrap procedure. Finally in \autoref{conclusion} we present and discuss our results.

\section{Review of the algorithm}\label{review}

All five-loop massless propagator integrals can be represented by families of the form
\begin{equation}\label{eq:prop-family}
	P_{k}(a_1,\ldots,a_{20};\DIM) \defas
	\int \frac{\td[\DIM] \LM_1}{\pi^{\DIM/2}}
	\cdots
	\int \frac{\td[\DIM] \LM_5}{\pi^{\DIM/2}}
	\ 
	\frac{1}{D_{1}^{a_1} \!\cdots D_{20}^{a_{20}}}
\end{equation}
where $\LM_i$ denote the loop momenta, and each $D_i$ is a quadratic form in these loop momenta and the external momentum $p$. There are 64 such families, indexed by $1\leq k \leq 64$, each corresponding to a different cubic graph with 14 internal lines, and $D_1,\ldots,D_{14}$ encode precisely the inverse propagators associated to these lines. The corresponding indices $a_1,\ldots,a_{14}$ will always be positive or zero. The remaining six quadratic forms $D_{15},\ldots,D_{20}$ will only appear with non-positive indices $a_{15},\ldots,a_{20} \leq 0$ and are needed to encode propagator integrals with numerators. See \autoref{max9-46-47} and \autoref{tab:D46}
 for the example of family $k=46$. We work in dimensional regularization with
$\DIM=4-2\dimeps$, so every propagator admits a Laurent expansion
\begin{equation}
	P(\DIM) = (p^2)^{-\sdc_{\dimeps}(P)} \sum_{n\in \ZZ} c_n(P)\dimeps^j
\label{eq:P-eps-expansion}%
\end{equation}
whose coefficients $c_n \in \RR$ are the numbers we want to determine.
Throughout we assume a positive definite (Euclidean) metric and we set $p^2=1$ in our calculations. The dependence on the external momentum $p$ is a power law and thus completely determined by the exponent
\begin{equation}\label{eq:sdc-eps}
	\sdc_{\dimeps} = \sum_{j=1}^{20} a_j - 5(2-\dimeps).
\end{equation}

\begin{table}\centering
\begin{tabular}{lllll}\toprule
$D_{1,\ldots,4}$ & $D_{5,\ldots,8}$ & $D_{9,\ldots,12}$ & $D_{13,\ldots,16}$ & $D_{17,\ldots,20}$
%$D_1,\ldots,D_4$ & $D_5,\ldots,D_8$ & $D_9,\ldots,D_{12}$ & $D_{13},\ldots,D_{16}$ & $D_{17},\ldots,D_{20}$
\\\midrule
$\LM_1^2$ & 
$\LM_3^2$ &
$(\LM_1 + \LM_3 + \LM_4 + \LM_5 - p)^2$ &
$(\LM_1 + \LM_4)^2$ &
$\LM_2\cdot \LM_4$ \\
$(\LM_1 - p)^2$ &
$\LM_4^2$ &
$(\LM_1 + \LM_3 + \LM_4 - p)^2$ &
$(\LM_2 -\LM_3 + p)^2$ &
$\LM_2 \cdot \LM_5$ \\
$\LM_2^2$ &
$(\LM_3 + \LM_4)^2$ &
$(\LM_2 - \LM_3 - \LM_5 + p)^2$ &
$\LM_1 \cdot \LM_3$ &
$\LM_3 \cdot \LM_5$ \\
$(\LM_2 + p)^2$ &
$\LM_5^2$ &
$(\LM_1 - \LM_2 + \LM_3 + \LM_4 + \LM_5 - p)^2$ &
$\LM_2\cdot \LM_3$ &
$\LM_4\cdot \LM_5$ \\
\bottomrule
\end{tabular}
\caption{Inverse propagators $D_{1},\ldots,D_{14}$ and numerators $D_{15},\ldots,D_{20}$ for integral family $P_{46}$.}%
\label{tab:D46}%
\end{table}

The algorithm of \cite{Baikov:2010hf} can be summarized in the following steps:
\begin{enumerate}
\item In each family $P_k$, choose a sufficient number of integer seeds $a \in \ZZ^{20}$ and reduce each of those p-integrals $P$ to (any preferred choice of) master integrals $M_i$:
\begin{equation}
	P(\DIM) = r_1(P,\DIM) M_1(\DIM) + \cdots + r_N(P,\DIM) M_N(\DIM)
\label{eq:P-M-expansion}%
\end{equation}
\item Expand both sides of \eqref{eq:P-M-expansion} in $\dimeps$ to express the Laurent coefficients $c_n(P)$ of each seed p-integral $P=P_k(a;\DIM)$ in terms of the corresponding coefficients $c_n(M_i)$ of the master integrals.
\item Enforce $c_n(P) = 0$ for all $n<-5$.
\item Enforce $c_n(P) = 0$ for all $n<0$ in the case of finite p-integrals.
\item Enforce the identities $c_0(P)=c_0(P')$ for all pairs of p-integrals $P$ and $P'$ that are related by the glue-and-cut symmetry.
\end{enumerate}

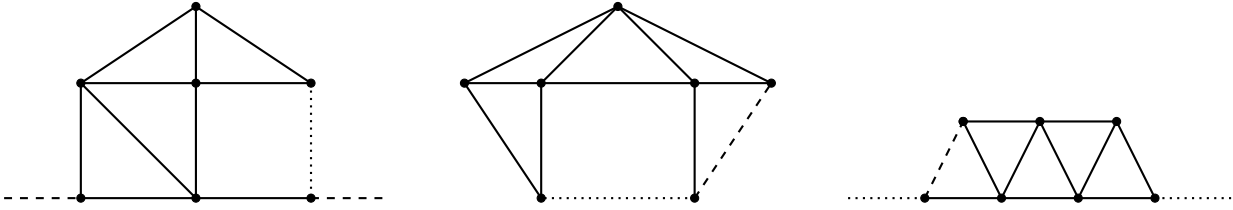
\begin{figure}
\begin{tikzpicture}[scale=0.51]
 \draw[dashed,thick] (0,0)--(2,0);
 \draw[dashed,thick] (8,0)--(10,0);
 \draw[thick] (2,0) -- (8,0);
 \draw[thick] (2,0)--(2,3)--(5,5)--(8,3);
 \draw[thick,dotted] (8,3)--(8,0);
 \draw[thick] (2,3)--(8,3);
 \draw[thick] (2,3)--(5,0)--(5,5);
 \draw[thick] (12,3)--(20,3);
 \draw[thick] (14,0)--(12,3)--(16,5)--(20,3);
 \draw[thick] (14,0)--(14,3)--(16,5)--(18,3)--(18,0);
 \draw[thick,dashed] (18,0)--(20,3);
 \draw[thick,dotted] (18,0)--(14,0);
 \draw[thick] (24,0)--(30,0);
 \draw[thick] (25,2)--(26,0)--(27,2)--(28,0)--(29,2)--(30,0);
 \draw[thick] (25,2)--(29,2);
 \draw[thick,dotted] (22,0)--(24,0);
 \draw[thick,dotted] (30,0)--(32,0);
 \draw[thick,dashed] (24,0)--(25,2);

  \filldraw[fill=black] (2,0) circle (3pt);
 \filldraw[fill=black] (2,3) circle (3pt);
 \filldraw[fill=black] (5,5) circle (3pt);
 \filldraw[fill=black] (8,3) circle (3pt);
 \filldraw[fill=black] (8,0) circle (3pt);
 \filldraw[fill=black] (5,0) circle (3pt);
 \filldraw[fill=black] (5,3) circle (3pt);

 \filldraw[fill=black] (14,0) circle (3pt);
 \filldraw[fill=black] (12,3) circle (3pt);
 \filldraw[fill=black] (16,5) circle (3pt);
 \filldraw[fill=black] (20,3) circle (3pt);
 \filldraw[fill=black] (14,3) circle (3pt);
 \filldraw[fill=black] (18,3) circle (3pt);
 \filldraw[fill=black] (18,0) circle (3pt);
 
 \filldraw[fill=black] (26,0) circle (3pt);
 \filldraw[fill=black] (27,2) circle (3pt);
 \filldraw[fill=black] (28,0) circle (3pt);
  \filldraw[fill=black] (29,2) circle (3pt);
 \filldraw[fill=black] (25,2) circle (3pt);
 \filldraw[fill=black] (30,0) circle (3pt);
 \filldraw[fill=black] (25,2) circle (3pt);
 \filldraw[fill=black] (24,0) circle (3pt);

\end{tikzpicture}
	\caption{Starting with the five-loop p-integral on the left and glueing the external lines (dashed), we obtain the six-loop vacuum diagram in the center. This diagram has 12 propagators and no numerator, leading to a vanishing superficial degree of divergence. Since there are no subdivergences, we can cut a different propagator (dotted) to produce a distinct five-loop p-integral with the same value in four dimensions.} \label{Cut}
\end{figure}

The glue-and-cut symmetry of massless p-integrals explains why sometimes different finite p-integrals evaluate to the same number in four dimensions. Diagrammatically, this process takes an $L$-loop p-integral $P$ and glues its two external legs together, to form an $(L+1)$-loop vacuum diagram $V$ (see \autoref{Cut} for an example).
Conversely, cutting the glued line from $V$, we recover $P$. But if we cut another line of $V$, we may produce a different p-integral $P'$. The glue-and-cut symmetry \cite{Baikov:2010hf} states that:
\begin{theorem}\label{thm:gluecut}
	If the p-integral $P(\dimeps)$ is finite in $\DIM=4$ dimensions ($\dimeps=0$), and if $V$ is dimensionless, then any other cut $P'(\dimeps)$ of $V$ is also finite in $\DIM=4$, and $P(0)=P'(0)$.
\end{theorem}
Dimensionless here means that the additional edge in $V$ is assigned the index $a_0 = \DIM/2-\omega_{\dimeps}(P)$, such that we have $\omega_{\dimeps}(V)=0$ where
\begin{equation*}
	\omega_{\dimeps}(V) \defas \sum_{j=0}^{20} a_j - 6(2-\dimeps).
\end{equation*}
In fact, as long as this condition is fulfilled, the symmetry holds for arbitrary values of the indices $a$ and for arbitrary $\dimeps$: $P(\dimeps)=P'(\dimeps)$ \cite{Panzer:MasslessPropagators}. Indeed, any $p$-integral $P$ obtained from cutting $V$, is equal to the residue of the vacuum integral $V$ at $\sdc_{\dimeps}(V)=0$.\footnote{%
	To make sense of a vacuum Feynman integral, it is in principle necessary to introduce some kinematic dependence, e.g.\ by adding some external legs, or assigning masses to some lines. However, the residue of $V$ at $\sdc=0$ does not depend on any of these choices.}
However, note that $a_0$ will depend on $\dimeps$, so in specializing to $\dimeps=0$ we gain that $a_0$ becomes an integer, and hence $P'(0)$ will again be a $p$-integral with integer indices.

This symmetry generates an array of identities between $c_0$-coefficients of different p-integrals. Once they are reduced to master integrals, we get a linear system of equations which in general allows us to relate the expansions of some master integrals in terms of the others. The hope is that the remaining undetermined coefficients can all be computed without difficulty.

The implementation of the algorithm for $L$-loop $p$-integrals starts then with the enumeration of $(L+1)$-loop vacuum diagrams $V$. Those are written as
\begin{equation*}
	\int \frac{\td[\DIM] \LM_1 \cdots \td[\DIM] \LM_{L+1}}{D_0^{a_0}\!\cdots D_n^{a_n}}\,,
\end{equation*}
with $D_i$ a linear combination of scalar products $\LM_i\cdot \LM_j$, which denotes either an inverse propagator ($a_i>0$) or a numerator ($a_i<0$). The superficial degree of divergence in four dimensions is then
\begin{equation}\label{sdc}
	\sdc_0 = \sum_{k=0}^n a_k - 2(L+1)\,.
\end{equation}
According to \autoref{thm:gluecut}, we consider only diagrams with $\sdc_0(V)=0$, which means that they must contain at least $2(L+1)$ propagators. The six-loop maximal sectors (diagrams with cubic vertices) have 15 denominators $D_0,\ldots,D_{14}$ (corresponding to the 15 edges in the graph), which means that in that case the candidates for cutting are obtained by adding $a_{15}+\ldots+a_{20}=3$ numerators. For graphs with less denominators ($a_i=0$ for one or more $i\leq 14$), we pick accordingly less numerators. Apart from thus assuring $\sdc_0(V)=0$, we need also to restrict to diagrams which have no subdivergences, so that the p-integrals resulting from cutting are all finite. Our algorithm for checking the convergence properties of vacuum diagrams will be explained in \autoref{convergence}.

\begin{figure}%[h]
\begin{tikzpicture}[scale=0.5]
 \draw[thick] (0,3)--(8,3);
 \draw[thick] (2,0)--(0,3)--(4,6)--(8,3);
 \draw[thick] (2,0)--(2,3)--(4,6)--(6,3)--(6,0);
 \draw[thick] (2,0)--(6,0)--(8,3);
 \filldraw[fill=white] (4,5.95) circle (5pt);
 \draw[thick] (18,0)--(20,3)--(18,6);
 \draw[thick] (14,6)--(12,3)--(14,0)--(18,0);
 \draw[thick] (12,3)--(20,3);
 \draw[thick] (14,0)--(14,6);
 \draw[thick] (18,0)--(18,6);
 \draw[thick,dotted] (18,6)--(14,6);
 \filldraw[fill=white] (14,5.95) circle (5pt);
 \filldraw[fill=white] (18,5.95) circle (5pt);
 \draw[thick] (24,3)--(30,3);
 \draw[thick] (24,3)--(24,5)--(26,3)--(26,5);
 \draw[thick] (28,5)--(28,3)--(30,5)--(30,3);
 \draw[thick] (24,5)--(30,5);
 \draw[thick,dotted] (22,3)--(24,3);
 \draw[thick,dotted] (30,3)--(32,3);
 \filldraw[fill=white] (24,3) circle (5pt);
 \filldraw[fill=white] (30,3) circle (5pt);

 \filldraw[fill=black] (0,3) circle (3pt);
 \filldraw[fill=black] (8,3) circle (3pt);
 \filldraw[fill=black] (2,0) circle (3pt);
 \filldraw[fill=black] (0,3) circle (3pt);
 \filldraw[fill=black] (6,0) circle (3pt);
 \filldraw[fill=black] (2,3) circle (3pt);
 \filldraw[fill=black] (6,3) circle (3pt);

  \filldraw[fill=black] (20,3) circle (3pt);
 \filldraw[fill=black] (18,0) circle (3pt);
 \filldraw[fill=black] (12,3) circle (3pt);
 \filldraw[fill=black] (14,0) circle (3pt);
 \filldraw[fill=black] (18,3) circle (3pt);
  \filldraw[fill=black] (14,3) circle (3pt);

 \filldraw[fill=black] (24,5) circle (3pt);
 \filldraw[fill=black] (26,3) circle (3pt);
 \filldraw[fill=black] (26,5) circle (3pt);
 \filldraw[fill=black] (28,5) circle (3pt);
  \filldraw[fill=black] (28,3) circle (3pt);
  \filldraw[fill=black] (30,5) circle (3pt);
 
\end{tikzpicture}
	\caption{We split the quartic vertex into a pair of cubic vertices connected by a new propagator (dashed). Cutting this propagator yields yet another finite p-integral.} \label{BlowUp} %
\end{figure}
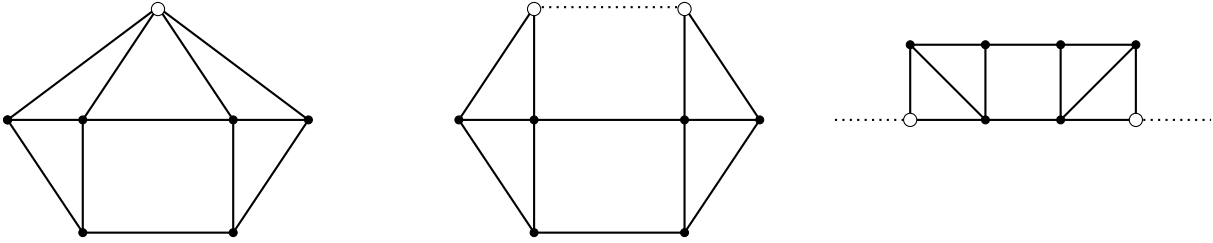

The following step is the cutting of the valid vacuum diagrams in all possible ways. The obvious identities are extracted by cutting along the existing denominators. However, one can also obtain constraints by splitting any of the vertices with valence $k>3$, see \autoref{BlowUp}. Starting for instance with a valence-$k$ vertex, we can produce two vertices of valences $(k-j+1)$ and $(j+1)$, as long as $j,(k-j)\geq 2$. These vertices are then connected by a new propagator (with index $a_i=0$) which we can cut, thus producing an additional identity.

Finally, the last step is the IBP reduction \eqref{eq:P-M-expansion} to master integrals. At
five loops this is quite an arduous task, further details on the
strategy followed are given in \autoref{reductions}. Once the
reduction is performed and all master integrals have been identified, one just has to perform the symbolic $\dimeps$ expansion \eqref{eq:P-eps-expansion}, impose finiteness of all p-integrals obtained from cutting, and equate their finite orders. Note that we also impose $c_n(P)=0$ for all $n<-5$, since even for divergent p-integrals, at five-loops they can at worst have poles of order five.

\section{Convergence}\label{convergence}
Consider a \emph{scalar} p-integral, that is, some $P_k(a;\DIM)$ as per \eqref{eq:prop-family} where $a_{15}=\cdots=a_{20}=0$ (no numerators). Such a p-integral is convergent if, and only if, we have $\sdc_{\dimeps}>0$ and the conditions
\begin{equation*}
	\sum_{e \in g} a_e
	-
	\frac{\DIM}{2} \cdot L_g
	>
	\begin{cases}
		\sdc_{\dimeps} & \text{if $g$ connects the external legs and} \\
		0      & \text{otherwise,} \\
	\end{cases}
\end{equation*}
are fulfilled for all non-empty proper subsets $g \subsetneq \{1,\ldots,14\}$ of the edges of the defining graph \cite{Weinberg:HighEnergy,Smirnov:AnalyticToolsFeynmanIntegrals}. Here, we denote by $L_g$ the loop number (first Betti number) of the subgraph $g$. Alternatively, consider the vacuum graph $V$ obtained by gluing the external legs of $P$ into an additional edge, with $a_0$ chosen such that $\sdc_{\dimeps}(V)=0$. Then the conditions above can equivalently be stated more symmetrically as
\begin{equation}
	\sum_{e \in g} a_e > \frac{\DIM}{2} \cdot L_g
	\label{eq:vacuum-finiteness}%
\end{equation}
for all non-empty subgraphs $g \subseteq\{0,\ldots,14\}$ of $V$. It is not necessary to check all of these conditions separately, however, as many of them turn out to be redundant. It suffices to verify the constraint only for those subgraphs $g$ that have the property that both $g$, and its quotient $V/g$, are biconnected \cite{Speer:SingularityStructureGenericFeynmanAmplitudes,Panzer:HeppBound}.

This provides an efficient way to select the finite integrals from a list of scalar p-integrals: For each of the 14 cubic graphs $V$ with six loops and 15 edges corresponding to a top-level vacuum integral, we compute once and for all the list of all subgraphs $g$ with the property that $g$ and $V/g$ are biconnected. This yields a short list (in the worst case there are $107$) of inequalities $\sum_{e \in g} a_e > \DIM/2 \cdot L_g$, which can be used to quickly determine if a scalar p-integral obtained by cutting some edge of $V$ is finite or not. For the remaining six-loop vacuum integrals (those with less than 15 edges), we proceed analogously.

With this control on scalar integrals, we can also determine the finiteness of integrals with numerators. Indeed, integrals with numerators can be viewed as linear combinations of scalar integrals in higher dimensions \cite{Tarasov:1996br}. Concretely, let us introduce Schwinger parameters $x_i$ for each denominator $D_i$ in \eqref{eq:prop-family}. Then, the quadratic form
\begin{equation*}
	x_1 D_1 + \cdots + x_{20} D_{20}
	= \LM^{\Transpose} A \LM + 2 B^{\Transpose} \LM + C
\end{equation*}
in the loop momenta $\LM=(\LM_{1},\ldots,\LM_{5})$ determines a $5\times 5$ matrix $A$, a 5-vector $B$ and a scalar $C$ (all of these depend on $x$). Define the polynomials $\U\defas \det A$ and $\F \defas \U(B^{\Transpose} A^{-1} B - C)$. The parametric representation \cite{Nakanishi:GraphTheoryFeynmanIntegrals,Smirnov:AnalyticToolsFeynmanIntegrals} of the integral family \eqref{eq:prop-family} then takes the form
\begin{equation}
	P_k(a;\DIM)
	= \Gamma(\sdc_{\dimeps})
	\left(
		\prod_{j\colon a_j >0} \int_0^{\infty} \frac{x_j^{a_j-1} \td x_j}{\Gamma(a_j)}
	\right)
	\left(
		\prod_{j\colon a_j \leq 0} \left[- \frac{\partial}{\partial x_j}\right]^{-a_j}_{x_j=0}
	\right)
	\frac{\delta(1-x_1)}{\U^{\DIM/2-\sdc_{\dimeps}}\F^{\sdc_{\dimeps}}}.
\label{eq:parametric-prop}%
\end{equation}
It is apparent that, after carrying out the derivatives with respect to $x_j$ for the numerators ($a_j<0$), and bringing the integrand on a common denominator, we can write $P_k(a;\DIM) = \sum_n b_n(\dimeps) P_k(a^{(n)};\DIM')$ as a finite linear combination of p-integrals in some raised dimension $\DIM'-\DIM \in 2 \ZZ_{\geq 0}$ that are scalar ($a^{(n)}_j=0$ whenever $a_j<0$), with raised indices $a_j^{(n)} \in a_j+\ZZ_{\geq 0}$ on the denominators ($a_j >0$).

The convergence of the integral \eqref{eq:parametric-prop} is equivalent to the convergence of each constituent $P_k(a^{(n)};\DIM')$.
We therefore apply the convergence criteria for scalar integrals as discussed above, in order to decide whether or not a given p-integral $P_k(a;\DIM)$ is finite, i.e.\ convergent in $\DIM=4$ dimensions. In fact, we carry out the entire analysis via \eqref{eq:vacuum-finiteness} on the level of the vacuum integrals. See \autoref{tab:finite-cuts} for statistics of the results. A detailed description will be given in \autoref{details}.
\begin{table}\centering
\begin{tabular}{ccccc}
\toprule
propagators & ISPs & numerators & graphs & finite integrals \\
\midrule
15 & 6 & 56 & 14 & 347 \\
14 & 7 & 28 & 22 & 248 \\
13 & 8 &  8 & 32 & 206 \\
12 & 9 &  1 & 17 &  15 \\
\bottomrule
\end{tabular}%
\caption{Counts of non-isomorphic vacuum graphs (see \autoref{Tadpoles}) by number of edges (propagators). The third column gives the number \eqref{eq:log-numerators} of logarithmically divergent numerators.}%
\label{tab:finite-cuts}%
\end{table}

The convergence of such an integral implies that $c_j(P_k(a;\DIM))=0$ for all singular coefficients ($j<0$) in the $\dimeps$-expansion \eqref{eq:P-eps-expansion}. Furthermore, we know also that the $c_0$ of such finite p-integrals agree for all cuts of the same vacuum integral. These constraints form the starting point of the {\GnC} algorithm.

\section{Reductions}\label{reductions}

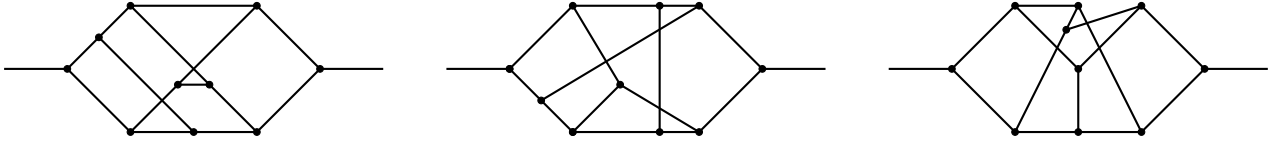
\begin{figure}
\centering
\begin{tikzpicture}[scale=0.42]
\draw[thick] (0,2)--(2,2);
\draw[thick] (10,2)--(12,2);
\draw[thick] (2,2)--(4,4)--(8,4)--(10,2)--(8,0)--(4,0)--(2,2);
\draw[thick] (4,0)--(8,4);
\draw[thick] (4,4)--(8,0);
\draw[thick] (6,0)--(3,3);
\draw[thick] (5.5,1.5)--(6.5,1.5);
\draw[thick] (14,2)--(16,2);
\draw[thick] (24,2)--(26,2);
\draw[thick] (16,2)--(18,4)--(22,4)--(24,2)--(22,0)--(18,0)--(16,2);
\draw[thick] (17,1)--(22,4);
\draw[thick] (18,0)--(19.5,1.5)--(22,0);
\draw[thick] (19.5,1.5)--(18,4);
\draw[thick] (20.75,4)--(20.75,0);

\draw[thick] (28,2)--(30,2);
\draw[thick] (38,2)--(40,2);
\draw[thick] (34,4)--(32,4)--(30,2)--(32,0)--(36,0)--(38,2)--(36,4);
\draw[thick] (32,0)--(34,4)--(36,0);
\draw[thick] (34,0)--(34,2)--(32,4);
\draw[thick] (34,2)--(36,4);
\draw[thick] (33.62,3.24)--(36,4);

 \filldraw[fill=black] (2,2) circle (3pt);
 \filldraw[fill=black] (4,4) circle (3pt);
 \filldraw[fill=black] (8,4) circle (3pt);
 \filldraw[fill=black] (10,2) circle (3pt);
 \filldraw[fill=black] (8,0) circle (3pt);
 \filldraw[fill=black] (4,0) circle (3pt);
 \filldraw[fill=black] (3,3) circle (3pt);
 \filldraw[fill=black] (6,0) circle (3pt);
 \filldraw[fill=black] (5.5,1.5) circle (3pt);
  \filldraw[fill=black] (6.5,1.5) circle (3pt);

\filldraw[fill=black] (16,2) circle (3pt);
 \filldraw[fill=black] (18,4) circle (3pt);
 \filldraw[fill=black] (22,4) circle (3pt);
 \filldraw[fill=black] (24,2) circle (3pt);
 \filldraw[fill=black] (22,0) circle (3pt);
 \filldraw[fill=black] (18,0) circle (3pt);
 \filldraw[fill=black] (16,2) circle (3pt);
 \filldraw[fill=black] (17,1) circle (3pt);
 \filldraw[fill=black] (18,0) circle (3pt);
 \filldraw[fill=black] (19.5,1.5) circle (3pt);
 \filldraw[fill=black] (20.75,4) circle (3pt);
 \filldraw[fill=black] (20.75,0) circle (3pt);

 \filldraw[fill=black] (30,2) circle (3pt);
 \filldraw[fill=black] (38,2) circle (3pt);
 \filldraw[fill=black] (34,4) circle (3pt);
 \filldraw[fill=black] (32,4) circle (3pt);
 \filldraw[fill=black] (32,0) circle (3pt);
 \filldraw[fill=black] (36,0) circle (3pt);
  \filldraw[fill=black] (34,2) circle (3pt);
  \filldraw[fill=black] (34,0) circle (3pt);
   \filldraw[fill=black] (36,4) circle (3pt);
 \filldraw[fill=black] (33.62,3.24) circle (3pt);
 
\end{tikzpicture}
\caption{The integral families for which bad denominators survive the change of basis. The diagram on the left shows family $P_{46}$.} \label{max9-46-47}%
\end{figure}

In order to apply the bootstrap procedure explained in \autoref{review}, we have to reduce the integrals involved to MI using IBP relations.
These relations are generated by setting to zero a total derivative,
\begin{equation}
	0=\int \frac{ \td[\DIM] \LM_1 }{\pi^{\DIM/2}} \cdots \frac{ \td[\DIM] \LM_L }{\pi^{\DIM/2}}
	\sum_{j=1}^L \frac{\partial}{\partial \LM_j^\mu}
	\frac{k_j^\mu \hspace{0.5mm} }{D_1^{a_1} \cdots D_n^{a_n}}\,,
\label{eq:IBP_schematic} %
\end{equation}
where $k_j$ is constructed as a linear combination of the external and internal momenta. With enough of these relations one can utilize the Laporta algorithm \cite{Laporta:2000dc} to row reduce the associated linear system.
There exist several available computer codes to perform such reductions \cite{Smirnov:2019qkx,Lee:2012cn,vonManteuffel:2012np,Klappert:2020nbg}.
In this work we applied the current version 
of {\FIRE}~\cite{Smirnov:2019qkx} (coupled with {\tt LiteRed}~\cite{Lee:2012cn})
with some features which are private at the moment but
will soon be made public, as usually. After revealing primary master integrals, we used the recently 
developed code~\cite{Smirnov:2020quc}, which can be used to improve a given basis of master integrals by getting rid of
so-called bad denominators from the coefficients of the IBP reduction\footnote{See also alternative competitive
code presented in~\cite{Usovitsch:2020jrk,Boehm:2020ijp}.}. By definition, a factor in the denominator is `good'
if it is either a linear function of $\DIM$ and independent of kinematic variables, or solely a polynomial of
the kinematic variables and thus independent of $\DIM$.
Any other configuration is considered `bad'.
Applying this code to the most complicated sectors we obtained a
new basis which is free of bad denominators in all but three of the
families\footnote{There could be other simpler sectors for which this
  happens but this simplification was not necessary to obtain the reduction.}, which are depicted in \autoref{max9-46-47}. 
Since the resolution of this issue is similar for all three families, we shall focus here only on the leftmost diagram of \autoref{max9-46-47}, which corresponds to family $P_{46}$ defined by \eqref{eq:prop-family} for the inverse propagators and numerators shown in \autoref{tab:D46}.

The fact that some bad denominators survive might be explained by the existence of a hidden 
relation between  master integrals. Unfortunately, the recipes based
on symmetries presented in~\cite{Smirnov:2013dia} did not help us find them.
In order to reveal this hidden relation we analyzed IBP reductions for a set of integrals with all but one of the exponents $a_i$ set to $0$ or $1$, while the remaining exponent is equal to $2$. By running {\FIRE} on integrals with thirteen positive indices in two different ways,
with the option {\tt no}$\_${\tt presolve} and without it, we then obtained two equivalent yet distinct reductions. By equating the corresponding results we then find the new relation, which has the form
\begin{multline}
	P_{46}(1,1,1,0,1,1,1,1,1,0,1,1,0,1,0,0,0,0,0,0;\DIM) \\
	\begin{aligned}
	=\frac{1}{3\DIM-11} \big[
		(8\DIM-28) & P_{46}(0,1,1,1,1,1,1,1,1,0,1,1,1,0,0,0,0,0,0,0;\DIM) \\
		-(5\DIM-25) & P_{46}(1,1,0,1,1,0,1,1,1,0,1,1,1,1,0,0,0,0,0,0;\DIM) \\
		+(\DIM- 5) & P_{46}(0,1,1,0,1,1,1,1,1,0,1,1,1,1,0,0,0,0,0,0;\DIM)
	\big] 
	+\ldots\quad
	\end{aligned}\label{eq:hidden_relation}%
\end{multline}
where dots stand for 37 master integrals with less than eleven positive indices.
The complete relation can be found in the ancillary file \texttt{ExtraRelation.m}.

After taking into account this additional relation and using the option \texttt{rules}
we observe that all the bad denominators disappear and the IBP reduction becomes faster
and requires less {\tt RAM}. The corresponding relations in the other two families of \autoref{max9-46-47} can be obtained from this one, since the relevant subsector is shared by all three families\footnote{Within {\FIRE}, one can use the function {\tt FindRules} for obtaining the mapping.}. 

By construction, this additional relation follows from IBP relations and symmetries
which come into the game with {\tt LiteRed}~\cite{Lee:2012cn} so that there is nothing 
mysterious in it. However, it looks reasonable to try to describe this and similar relations
in a more natural language.

\section{Details of our calculation}\label{details}

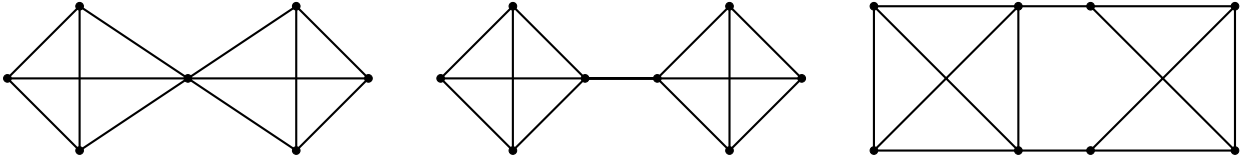
\begin{figure}[t]
\begin{tikzpicture}[scale=0.48]
\draw[thick] (0,2)--(10,2);
\draw[thick] (0,2)--(2,4)--(5,2)--(8,4)--(10,2)--(8,0)--(5,2)--(2,0)--(0,2);
\draw[thick] (2,4)--(2,0);
\draw[thick] (8,0)--(8,4);
\draw[thick] (12,2)--(22,2);
\draw[thick] (12,2)--(14,4)--(16,2)--(18,2)--(20,4)--(22,2)--(20,0)--(18,2)--(16,2)--(14,0)--(12,2);
\draw[thick] (14,4)--(14,0);
\draw[thick] (20,0)--(20,4);
\draw[thick] (24,4)--(34,4);
\draw[thick] (24,0)--(34,0);
\draw[thick] (24,0)--(24,4)--(28,0)--(28,4)--(24,0);
\draw[thick] (30,4)--(34,0)--(34,4)--(30,0);

\filldraw[fill=black] (0,2) circle (3pt);
\filldraw[fill=black] (2,4) circle (3pt);
\filldraw[fill=black] (5,2) circle (3pt);
\filldraw[fill=black] (8,4) circle (3pt);
\filldraw[fill=black] (10,2) circle (3pt);
\filldraw[fill=black] (8,0) circle (3pt);
\filldraw[fill=black] (2,0) circle (3pt);

\filldraw[fill=black] (12,2) circle (3pt);
\filldraw[fill=black] (14,4) circle (3pt);
\filldraw[fill=black] (16,2) circle (3pt);
\filldraw[fill=black] (18,2) circle (3pt);
\filldraw[fill=black] (20,4) circle (3pt);
\filldraw[fill=black] (22,2) circle (3pt);
\filldraw[fill=black] (20,0) circle (3pt);
\filldraw[fill=black] (14,0) circle (3pt);

\filldraw[fill=black] (24,4) circle (3pt);
\filldraw[fill=black] (34,4) circle (3pt);
\filldraw[fill=black] (24,0) circle (3pt);
\filldraw[fill=black] (34,0) circle (3pt);
\filldraw[fill=black] (28,0) circle (3pt);
\filldraw[fill=black] (28,4) circle (3pt);
\filldraw[fill=black] (30,4) circle (3pt);
\filldraw[fill=black] (30,0) circle (3pt);

\end{tikzpicture}
	\caption{The diagrams on the left and center are rejected since they would produce p-integrals with tadpoles after the cutting procedure. The condition for removal is the existence of a zero-momentum edge (manifest in the center diagram but revealed on the left only after blow-up of the higher-valence vertex). Meanwhile the rightmost diagram is rejected because it produces integrals with a double propagator.} \label{Invalid}
\end{figure}
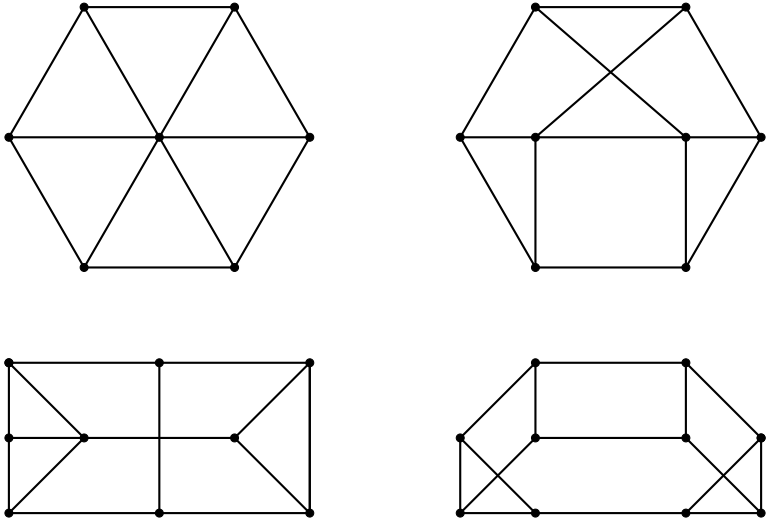
\begin{figure}[t]
\centering
\begin{tikzpicture}[scale=0.5]
\draw[thick] (0,3)--(8,3)--(6,3+4*0.866)--(2,3+4*0.866)--(0,3)--(2,3-4*0.866)--(6,3-4*0.866)--(8,3);
\draw[thick] (6,3+4*0.866)--(2,3-4*0.866);
\draw[thick] (6,3-4*0.866)--(2,3+4*0.866);
\draw[thick] (12,3)--(20,3)--(18,3+4*0.866)--(14,3+4*0.866)--(12,3)--(14,3-4*0.866)--(18,3-4*0.866)--(20,3);
\draw[thick] (14,3-4*0.866)--(14,3)--(18,3+4*0.866);
\draw[thick] (18,3-4*0.866)--(18,3)--(14,3+4*0.866);
\draw[thick] (0,-7)--(8,-7);
\draw[thick] (0,-7)--(0,-3)--(8,-3)--(8,-7);
\draw[thick] (0,-5)--(6,-5)--(8,-3)--(8,-7)--(6,-5);
\draw[thick] (0,-3)--(2,-5)--(0,-7);
\draw[thick] (4,-7)--(4,-3);
\draw[thick] (12,-7)--(20,-7);
\draw[thick] (20,-7)--(20,-5)--(18,-7);
\draw[thick] (12,-7)--(12,-5)--(14,-7);
\draw[thick] (12,-5)--(14,-3)--(18,-3)--(20,-5);
\draw[thick] (14,-3)--(14,-5)--(18,-5)--(18,-3);
\draw[thick] (12,-7)--(14,-5);
\draw[thick] (20,-7)--(18,-5);

\filldraw[fill=black] (0,3) circle (3pt);
\filldraw[fill=black] (8,3) circle (3pt);
\filldraw[fill=black] (6,3+4*0.866) circle (3pt);
\filldraw[fill=black] (2,3+4*0.866) circle (3pt);
\filldraw[fill=black] (2,3-4*0.866) circle (3pt);
\filldraw[fill=black] (6,3-4*0.866) circle (3pt);
\filldraw[fill=black] (4,3) circle (3pt);

\filldraw[fill=black] (12,3) circle (3pt);
\filldraw[fill=black] (20,3) circle (3pt);
\filldraw[fill=black] (18,3+4*0.866) circle (3pt);
\filldraw[fill=black] (14,3+4*0.866) circle (3pt);
\filldraw[fill=black] (14,3-4*0.866) circle (3pt);
\filldraw[fill=black] (18,3-4*0.866) circle (3pt);
\filldraw[fill=black] (14,3) circle (3pt);
\filldraw[fill=black] (18,3) circle (3pt);

\filldraw[fill=black] (0,-7) circle (3pt);
\filldraw[fill=black] (8,-7) circle (3pt);
\filldraw[fill=black] (0,-3) circle (3pt);
\filldraw[fill=black] (8,-3) circle (3pt);
\filldraw[fill=black] (6,-5) circle (3pt);
\filldraw[fill=black] (0,-3) circle (3pt);
\filldraw[fill=black] (2,-5) circle (3pt);
\filldraw[fill=black] (0,-5) circle (3pt);
\filldraw[fill=black] (4,-3) circle (3pt);
\filldraw[fill=black] (4,-7) circle (3pt);

\filldraw[fill=black] (12,-7) circle (3pt);
\filldraw[fill=black] (12,-5) circle (3pt);
\filldraw[fill=black] (14,-7) circle (3pt);
\filldraw[fill=black] (14,-3) circle (3pt);
\filldraw[fill=black] (18,-3) circle (3pt);
\filldraw[fill=black] (20,-5) circle (3pt);
\filldraw[fill=black] (18,-5) circle (3pt);
\filldraw[fill=black] (20,-7) circle (3pt);
\filldraw[fill=black] (20,-5) circle (3pt);
\filldraw[fill=black] (18,-7) circle (3pt);
\filldraw[fill=black] (14,-5) circle (3pt);

\end{tikzpicture}
	\caption{Examples of vacuum diagrams at six loops with $12$, $13$, $14$ and $15$ denominators (reading from left to right and top to bottom). The larger the number of propagators, the larger the number of numerators we must include to ensure vanishing $\sdc_0$.} \label{Tadpoles} %
\end{figure}

Using the computer program {\tt SAGE}\cite{sagemath}, we found 102 six-loop vacuum diagrams with at least 12 propagators, 99 of which are connected. Removing graphs with zero-momentum edges or double propagators (see Figure~\ref{Invalid} for more details), we are left with 85 diagrams which are viable for cutting, some of which are depicted in Figure~\ref{Tadpoles}.

In order to have zero superficial degree of divergence $\sdc_0$,
defined in \eqref{sdc}, the diagrams with $(12+x)$ propagators must
be accompanied by $x$ numerators. The complete basis of scalar
products has 21 elements at this loop order, therefore the numerator
can be any polynomial of degree $x$ built from the $(9-x)$ irreducible
scalar products (ISPs) available. In practice we avoided the complex search for all generic convergent numerators, and instead we only tested convergence for each monomial. We have many available integrals nevertheless, since each diagram with $(12+x)$ propagators produces 
\begin{equation}
(9-x)_x/x!
\label{eq:log-numerators}%
\end{equation}
distinct numerators, with $(a)_x$ the Pochhammer symbol. However, since the IBP reductions are very demanding at this loop order, we neglected all vacuum diagrams with $15$ denominators and degree $3$ numerators, and checked only the convergence properties of the remaining $889 = 28\cdot 22 + 8\cdot 32 + 1\cdot 17$ integrals in \autoref{tab:finite-cuts}, finding that $469 = 248+206+15$ of them are free of divergences.

The attentive reader will wonder how it is possible to fix the expansions of master p-integrals with $14$ denominators if we cut only vacuum diagrams with $14$ denominators or less. The trick resides in the splitting of higher-valence vertices, which effectively increases the number of propagators by one.
Note also that we checked convergence for numerators which are written as products of ISPs, hence the number of finite integrals in \autoref{tab:finite-cuts} depend on our choice of ISPs. We considered two options, a difference basis given by $(\LM_i-\LM_j)^2$, and a dot basis formed by $\LM_i \cdot \LM_j$ (for example see $D_{15},\ldots,D_{20}$ in \autoref{tab:D46}), with $\LM_1,\ldots,\LM_5$ the loop momenta. The analysis at lower loops showed that the latter produced more convergent integrals, and therefore more equations, and so we considered the dot basis for the bootstrap at five loops.

We then systematically cut each of the $469$ finite vacuum diagrams in all possible ways, either through an existing propagator, or by splitting a higher-valence vertex, as explained in \autoref{review}. In the end we obtained $7647$ equations relating different cuttings of the vacuum diagrams. Each cut produces a single finite p-integral with a given numerator which is inherited from that of the vacuum diagram, but one needs to embed the integral into one of the $64$ five-loop maximal sectors (diagrams with cubic vertices). Therefore, one must decompose the numerator into the basis of the chosen maximal sector, which effectively produces a linear combination of many divergent p-integrals that need to be reduced.

Once we run the IBP reductions for each of the $64$ families, we
obtain a set of master integrals for each of them. However, some of
the integrals are common to several families. For example, the
watermelon diagram inevitably appears in the reductions of all $64$
maximal sectors. While it is easy to find relations between subsectors of
different families, one must be careful when a given
sector has more than one master integral. The reduction routine
outputs in that case master integrals with a double denominator, but
the precise prescription for its positioning
depends on the family upon consideration. In order to avoid
redundancies one must therefore set a convention for the sector that
provides the preferred integral, and then map to it all variants arising from other families with a different double-denominator position.

After all such considerations, we find that at five loops there are
$281$ master integrals from $245$ distinct sectors.
There are $30$
sectors with $2$ master integrals, and also $3$ sectors containing $3$
master integrals each. Only $16$ of the integral families $P_k$ from \eqref{eq:prop-family} contain master integrals in the top sector (that means $a_i>0$ for \emph{all} propagators $1\leq i \leq 14$), but $12$ of these families do contain more than one master integral in the top sector.

In principle it would have been possible to miss some master integrals, since the maximal sectors of a few integral families are not probed by any of the integrals produced through the cutting procedure. However, we have used \Azurite~\cite{Georgoudis:2016wff} to confirm that there are no extra master integrals in any of those sectors. Furthermore, and since we applied the glue-and-cut algorithm to integrals with at most $2$ numerators, we have also used {\Azurite} to reduce all maximal sectors with degree $3$ and verified the absence of new master integrals. 
 
It is clear that the linear system of equations will impose a myriad of
relations between the expansions of the different master integrals. However, the key point is whether we can fix all expansions in terms of coefficients that we can easily evaluate.
At four loops this turned out to be possible, with all master integrals given in terms of the expansions of recursively one-loop integrals. This class of diagrams can be evaluated exactly by a recursive application of the one-loop bubble integration
\begin{equation}\label{ROL}
\frac{1}{\dimeps G(1,1)} \int\frac{\td[\DIM] \LM}{\pi^{\DIM/2}} \frac{1}{\LM^{2a} (p-\LM)^{2b}} = (p^2)^{\DIM/2-a-b} \frac{G(a,b)}{\dimeps \,G(1,1)}\,,
\end{equation}
with $G$ defined as
\begin{equation}\label{G}
G(a,b)= \frac{\Gamma(a+b-\DIM/2) \Gamma(\DIM/2-a) \Gamma(\DIM/2-b) }{ \Gamma(a) \Gamma(b) \Gamma(\DIM-a-b)}\,.
\end{equation}
Note that we used the freedom in the definition of dimensional regularization to normalize integrations according to the $G$-scheme \cite{Chetyrkin:1980pr}. Using the functional equation $\Gamma(a+1)=a \Gamma(a)$, the $\dimeps$-expansion of $G(a,b)/G(1,1)$ consists, at any order, of rational linear combinations of Riemann zeta values: For example, near $a=1$, the expansion can be obtained from
\begin{equation}
	\frac{\Gamma(2-\dimeps-a)}{\Gamma(a)} \Big/ \frac{\Gamma(1-\dimeps)}{\Gamma(1)}
	=\exp\left[ \sum_{k=2}^{\infty} \frac{\zeta(k)}{k}\Big(  
		(\dimeps+a-1)^n-(1-a)^n-\dimeps^n
	\Big) \right].
	\label{eq:Gamma-exp}%
\end{equation}

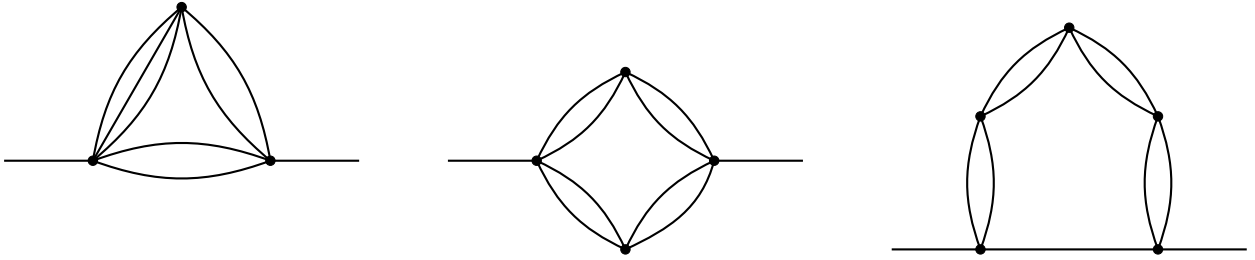
\begin{figure}
\begin{tikzpicture}[scale=0.59]
 \draw[thick] (0,2)--(2,2);
 \draw[thick] (6,2)--(8,2);
 \draw[thick] (2,2)--(4,2+4*0.866);
 \draw[thick]  (2,2)to[out=80,in=220](4, 2+4*0.866);
 \draw[thick]  (4, 2+4*0.866)to[out=-100,in=40](2, 2);
 \draw[thick]  (2,2)to[out=20,in=160](6, 2);
 \draw[thick]  (6,2)to[out=200,in=-20](2, 2);
 \draw[thick]  (6,2)to[out=140,in=-80](4, 2+4*0.866);
 \draw[thick]  (4, 2+4*0.866)to[out=-40,in=100](6, 2);
 \draw[thick] (10,2)--(12,2);
 \draw[thick] (16,2)--(18,2);
 \draw[thick]  (12,2)to[out=65,in=205](14, 4);
 \draw[thick]  (14,4)to[out=245,in=25](12,2);
 \draw[thick]  (12,2)to[out=-25,in=115](14, 0);
 \draw[thick]  (14,0)to[out=155,in=-65](12,2);
 \draw[thick]  (16,2)to[out=205,in=65](14, 0);
 \draw[thick]  (14,0)to[out=25,in=255](16,2);
  \draw[thick]  (16,2)to[out=115,in=-25](14, 4);
 \draw[thick]  (14,4)to[out=-65,in=155](16,2);
 \draw[thick]  (20,0)--(28,0);
 %\draw[thick]  (22,0)--(22,3)--(24,5)--(26,3)--(26,0);
  \draw[thick]  (22,0)to[out=110,in=250](22, 3);
 \draw[thick]  (22,3)to[out=-70,in=70](22,0);
  \draw[thick]  (26,0)to[out=110,in=250](26, 3);
 \draw[thick]  (26,3)to[out=-70,in=70](26,0);
  \draw[thick]  (22,3)to[out=65,in=205](24, 5);
 \draw[thick]  (24,5)to[out=245,in=25](22,3);
   \draw[thick]  (26,3)to[out=115,in=-25](24,5);
   \draw[thick]  (24,5)to[out=-65,in=155](26,3);

   \filldraw[fill=black] (2,2) circle (3pt);
   \filldraw[fill=black] (4,2+4*0.866) circle (3pt);
   \filldraw[fill=black] (6,2) circle (3pt);

   \filldraw[fill=black] (12,2) circle (3pt);
   \filldraw[fill=black] (16,2) circle (3pt);
   \filldraw[fill=black] (14,4) circle (3pt);
   \filldraw[fill=black] (14,0) circle (3pt);

   \filldraw[fill=black] (22,0) circle (3pt);
   \filldraw[fill=black] (22,3) circle (3pt);
   \filldraw[fill=black] (24,5) circle (3pt);
   \filldraw[fill=black] (26,3) circle (3pt);
   \filldraw[fill=black] (26,0) circle (3pt);

\end{tikzpicture}
	\caption{Three examples of recursively one-loop master integrals. Each loop variable can be integrated one by one, and in the end we obtain an exact expression for each of these integrals as a function of the spacetime dimension $\DIM$.} \label{Rol}%
\end{figure}

Meanwhile, at five loops there are $21$ recursively one-loop master integrals appearing in the IBP reductions of p-integrals (see \autoref{Rol} for some examples). Equation~\eqref{ROL} can then be used to obtain their expansions explicitly, so that we provide some inputs to the linear system of equations. In our case, however, these integrals cannot provide sufficient information to fix all expansions up to transcendental weight $9$. Namely, at five loops there is for the first time a contribution from a multiple zeta value (MZV) of weight $8$. MZVs are defined as
\begin{equation}
	\mzv{n_1,\ldots,n_r} = \sum_{1\leq k_1 < \cdots < k_r} \frac{1}{k_1^{n_1} \!\cdots k_r^{n_r}}
	\in \RR_{>0},
	\label{eq:MZV}%
\end{equation}
where the indices $n_1,\ldots,n_{r-1}\geq 1$ and $n_r \geq 2$ denote integers, and their sum $n_1+\cdots+n_r$ is referred to as the transcendental \emph{weight}. Up to weight $7$, all such MZV can be written as rational linear combinations of products of Riemann zeta values ($r$=1). But at weight $8$, such relations still leave a $1$-dimensional quotient space, generated e.g.\ by $\mzv{3,5}$, that conjecturally cannot be reduced in this way, and can hence never be obtained from the expansion of \eqref{G}, since the latter only contains Riemann zeta values from \eqref{eq:Gamma-exp}. As $\mzv{3,5}$ has long been known to appear at five-loop order in massless propagators \cite{Broadhurst:5loopsbeyond,Broadhurst:1440}, it is thus clear that we will need further inputs, beyond recursively one-loop integrals.

At this point the linear system of equations obtained from the cutting procedure has been written in terms
of the master integrals. We input the expansions of all recursively one-loop integrals, which we can evaluate explicitly according to equation \eqref{ROL}. However, before we consider the relations introduced by different cuttings, we can already
extract some constraints from an analysis of the IBP identities
themselves. While five-loop p-integrals diverge at most as $\dimeps^{-5}$, one finds that some of the coefficients $r_i$ in the reductions \eqref{eq:P-M-expansion} also contain poles in $\dimeps$. Requiring
the cancellation of those (spurious) poles of order $\dimeps^{-6}$ and worse, fixes already $563$ coefficients in
the expansions of the master integrals.
Then we impose that all  $8116$  p-integrals obtained from distinct cuttings of the finite vacuum diagrams are finite, which further fixes $1300$ coefficients. Finally, looking at
the $7647$ equations which relate the finite orders of those p-integrals,
we are able to fix another $259$ coefficients. 

In total we were able to fix $2122$ coefficients using the expansions
of the $21$ recursively one-loop integrals, which themselves contain
only $264$ coefficients. In the end, the expansions of all master
integrals up to transcendental weight 9 depend only on $2$ undetermined coefficients. As explained above, they encode the dependence on the expected MZV of weight $8$.  Fortunately, these undetermined coefficients are present in the expansions of many integrals, including $21$ which are products of lower-loop masters. All lower-loop master integrals were computed up to transcendental weight twelve\footnote{At four loops, this was done in \cite{Lee:2011jt}.}, and so we can easily determine the expansions of all five-loop master integrals which are a product of lower-loop integrals.

\begin{figure}
\centering
\begin{tikzpicture}[scale=1]
\draw[thick] (0,0)--(4,0);
\draw[thick]  (1.5,0)to[out=20,in=160](4,0);
\draw[thick]  (4,0)to[out=200,in=-20](1.5,0);
\draw[thick] (4,0)--(6,1)--(8,0)--(6,-1)--(4,0);
\draw[thick]  (6,-1)to[out=110,in=250](6,1);
\draw[thick]  (6,1)to[out=-70,in=70](6,-1);
\draw[thick] (8,0)--(9.5,0);

\filldraw[fill=black] (1.5,0) circle (2pt);
\filldraw[fill=black] (4,0) circle (2pt);
\filldraw[fill=black] (6,1) circle (2pt);
\filldraw[fill=black] (8,0) circle (2pt);
\filldraw[fill=black] (6,-1) circle (2pt);

\end{tikzpicture}
	\caption{This five-loop master integral is a product of two-and three-loop master integrals. The knowledge of its expansion, together with the recursively one-loop integrals, is sufficient to determine the expansions of all remaining master integrals.} \label{Extra} %
\end{figure}
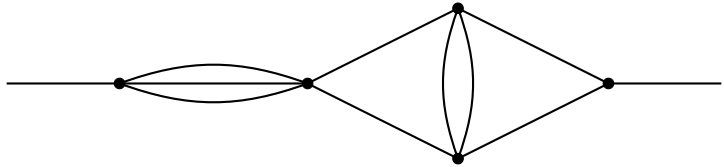

In fact, the last $2$ undetermined coefficients can be extracted from the expansion of the p-integral depicted in \autoref{Extra}. Using \eqref{ROL}, the missing information is thus contained in known higher order corrections to the two-loop master integral \cite{Broadhurst:1440,Bierenbaum:2003ud}.

\section{Results}
\label{conclusion}

In this work we have demonstrated that the glue-and-cut algorithm
still provides an extremely powerful way of determining master
p-integrals at the five-loop order. Following the strategy developed
at lower loops, we input the expansions of $21$ recursively one-loop
integrals, which can be evaluated easily to arbitrary orders in
$\varepsilon$. The system of equations obtained fixes the expansions
of the remaining $260$ master-integrals, up to transcendental weight
$9$, in terms of a single master integral, which can be chosen as the
product of planar lower-loop diagrams.

As we have mentioned in \autoref{details}, with our choice of master integrals,\footnote{Complete details are provided in the supplementary files.} some coefficients $r_i(P,4-2\dimeps)$ of the IBP reductions \eqref{eq:P-M-expansion} occasionally have poles at $\dimeps=0$. These poles are called \emph{spurious poles} in \cite{Baikov:2010hf}. Such poles do not necessarily cause a problem, since the {\GnC} constraints determine each master integral $M_i$ all the way up to the order $q_i$ where terms of transcendental weight 9 first appear. However, and unlike the four-loop case, we observe instances where the maximal power of a spurious pole is greater than the order $q_i$ to which the master integral is fixed. While these higher orders are not fully determined, the {\GnC} constraints are still sufficient because we find that all remaining degrees of freedom arising from higher order coefficients $c_{>q_i}(M_i)$ effectively cancel out from the coefficients $c_{\leq 0}(P)$ of any p-integral $P$.

Furthermore, and since we are free to choose any basis of master integrals, we followed the procedure indicated in \cite{Chetyrkin:2006dh} and constructed a new basis such that the IBP reduction coefficients never exhibit poles in $\dimeps$ with degree higher than $q_i$. This new basis is obtained by selecting a master integral $M_i$ for which the power of the spurious poles surpasses $q_i$, and replacing it with one of the $p$-integrals $P$ whose coefficient $r_i(P,\DIM)$ realizes the maximal pole. This procedure is iterated, until all such spurious poles are resolved.\footnote{It is interesting to notice that the number of replaced master integrals is considerably less than the number of integrals with undetermined higher-order coefficients, which points to a common structure of the spurious divergent poles in a specific dimension.} 
The basis thus constructed, and its relation to our original basis, is provided in the supplementary files. 

We checked that the expansions of the planar integrals precisely match the results obtained in the context of position-space integrals \cite{Georgoudis:2018olj}. Similarly, the expansions of all non-planar product-type integrals agree with those obtained from lower-loop master integrals. Finally, a few orders in the expansions of some genuine five-loop non-planar integrals which were known from the Fourier transform of position-space results \cite{Georgoudis:2018olj} and a $\phi^4$ computation \cite{Kompaniets:2017yct}, were all equally confirmed here. For the finite p-integrals obtained as cuts of $\phi^4$ four-point graphs, our results agree with the periods determined in \cite{Schnetz:2008mp}.

We wish also to emphasize that the equations obtained with this method do not discriminate between planar and non-planar graphs, and they often involve p-integrals from both categories. In that way, it seems highly unlikely that errors could arise only for the non-planar graphs for which we provide novel results.

Interestingly, we find that the $\dimeps$-expansions of all master integrals up to transcendental weight 9, and moreover all coefficients $c_n(P)$ with $n\leq 0$ of all p-integrals we considered, can be written as polynomials (with rational coefficients) in the following five quantities:
\begin{equation}\label{eq:mzveps}\begin{split}
\mzveps{3} &= \mzv{3} + \frac{3 \dimeps}{2}\mzv{4} - \frac{5 \dimeps^3}{2} \mzv{6} + \frac{21 \dimeps^5}{2} \mzv{8}\,,  \qquad 
\mzveps{5}  = \mzv{5} + \frac{5 \dimeps}{2} \mzv{6} - \frac{35 \dimeps^3}{4} \mzv{8} \,,\\
\mzveps{7} &= \mzv{7} + \frac{7 \dimeps}{2} \mzv{8} \,, \qquad
\mzveps{3,5}  = \mzv{3,5} - \frac{29}{12} \mzv{8} -\frac{15\dimeps}{2} \mzv{4} \mzv{5}, \quad\text{and}\quad
\mzv{9} \,.
\end{split}\end{equation}
The same observation and $\dimeps$-dependent transformation was found previously in the context of position space integrals \cite{Georgoudis:2018olj}, up to a redefinition of $\mzveps{3,5}$ following \cite{Baikov:2019zmy} that is purely a matter of convenience.%
\footnote{
	In terms of $\mzveps{3,5}' = \mzv{2,6}- \mzv{5,3} - 3 \dimeps \mzv{4} \mzv{5} + \frac{5 \dimeps}{2} \mzv{3} \mzv{6}$ as defined in \cite{Georgoudis:2018olj}, the precise relation is
	 $\mzveps{3,5}' = \frac{3}{5}\mzveps{3,5} +\mzveps{3} \mzveps{5}$.
}
This implies that, up to the $\dimeps$-order where the master integrals reach transcendental weight 9, the contributions involving even zeta values are not independent, but in fact determined completely by the coefficients (in lower $\dimeps$-orders) of polynomials in
\begin{equation}\label{eq:mzv0}
\mzv{3}\,, \quad \mzv{5}\,,\quad \mzv{7}\,,\quad \mzv{3,5}- \frac{29}{12}\mzv{8} \,,\quad\text{and}\quad \mzv{9}\,,
\end{equation}
The first such connection, tying $\mzv{4}=\frac{\pi^4}{90}$ to
$\mzv{3}$ at three loops, was explored in
\cite{Broadhurst:DimContGauge}. At the four-loop level, the
correlation of $\mzv{6}=\frac{\pi^6}{945}$ to $\mzv{3}$ and $\mzv{5}$
was settled in \cite{Baikov:2010hf}, and extensions of such relations
and $\dimeps$-dependent transformations like \eqref{eq:mzveps} up to
eight loops (transcendental weight 13) were proposed in
\cite{Baikov:2019zmy}. Our work can be viewed, alongside
\cite{Georgoudis:2018olj}, as a further indication towards the
validity of this picture at five loops (transcendental weight 9),
where multiple zeta values enter for the first time, as $\mzveps{3,5}$.
These findings suggest that the rich number-theoretic structure \cite{PanzerSchnetz:Phi4Coaction} of \emph{convergent} ($\dimeps=0$) p-integrals persists also in some form on the level of $\dimeps$-expansions, which is a domain whose mathematical foundations are still under construction \cite{BrownDupont:Lauricella}.
A systematic understanding of these $\dimeps$-dependent transformations of transcendentals has so far been obtained only in the case of Riemann zeta values \cite{KotikovTeber:LKFzeta}.

These constraints on the $\dimeps$-expansions of p-integrals have striking implications for renormalization group functions, in form of the \emph{no-$\pi$ theorem} \cite{BaikovChetyrkin:nopi}, which predicts the $\pi$-dependent terms (involving even zeta values) in terms of the $\pi$-free terms.

Given its success so far, it seems reasonable to expect that the glue-and-cut method should in principle be applicable also at six loops. However, it requires the ability to reduce p-integrals to a basis of master integrals, and at six loops that task does not seem feasible with any of the currently available tools for IBP reduction.

Finally, the p-integrals presented in this paper have different
applications, beside the aforementioned computation of $\beta$-functions or anomalous dimensions. For example, they can also appear as boundary conditions in the context of differential
equations when computing higher-point integrals \cite{Henn:2016men,Eden:2016dir,Bruser:2020bsh}. 
  
\section*{Acknowledgments}
%%%%%%%%%%%%%%%%%%%%%%%%%%%%%%%%%%%%%%%%%%%%%%%%%%%%%%%%%
The authors would like to thank Kostja Chetyrkin for useful discussions
regarding the glue-and-cut method.
AG also thanks Tiziano Peraro for providing the reduction of one of
the families of p-integrals, using the method of \cite{Peraro:2019svx}. 
Part of the computations were performed on resources provided by the Swedish National Infrastructure for Computing (SNIC)  at Uppmax.
V.~G.\ is funded by FAPESP grant 2015/14796-7 ,
CERN/FIS-NUC/0045/2015 and Simons Foundation grant 488637 (Simons Collaboration on the Non-perturbative bootstrap). The work of A.~G.\ is supported by the Knut and
Alice Wallenberg Foundation under grant 2015.0083 and by the French
National Agency for Research grant ANR-17-CE31-0001-02. A.~G.\  would
like to thank FAPESP grant 2016/01343-7 for funding part of his visit
to ICTP-SAIFR in March 2019 where part of this work was done. R.~P.\ was
funded under SFI grant 15/CDA/3472. The work of A.~S.\ and V.~S.\ was supported by Russian Ministry of Science and Higher Education, agreement No. 075-15-2019-1621.
E.~P.\ is funded as a Royal Society University Research Fellow through grant {URF{\textbackslash}R1{\textbackslash}201473}.
A.~G.\ and V.~S.\ are grateful to the organizers of the Paris Winter Workshop: The Infrared in QFT (March 2020)
where we understood how the present project could be completed.

\bibliographystyle{JHEP}
\bibliography{biblio}

\end{document}